# Quantitative topographic imaging using a near-field scanning microwave microscope

C. P. Vlahacos, D. E. Steinhauer, S. K. Dutta, B. J. Feenstra, Steven M. Anlage, and

F. C. Wellstood

*Center for Superconductivity Research*

*Department of Physics*

*University of Maryland*

*College Park, MD 20742-4111*

We describe a technique for extracting topographic information using a scanning near-field microwave microscope. By monitoring the shift of the system's resonant frequency, we obtain quantitative topographic images of uniformly conducting metal surfaces. At a frequency of 9.572 GHz, our technique allows for a height discrimination of about 55 nm at a separation of 30 µm. We present topographic images of uneven, conducting samples and compare the height response and sensitivity of the system with theoretical expectations.

Images obtained from resonant near-field scanning microwave microscopes[1-5] are the result of two distinct contrast mechanisms: shifts in the resonant frequency due to electrical coupling between the probe and the sample and changes in the quality factor Q of the resonator due to losses in the sample.[4,5] Such images will inevitably contain intrinsic information about a sample (such as dielectric constant or surface resistance[5]) as well as extrinsic information (such as surface topography[6]). To facilitate quantitative imaging of intrinsic material properties, it is essential to be able to accurately account for the effects of finite probe-sample separation and topography.[6,7] In this Letter, we report on the extraction of calibrated topographic images using a resonant near-field scanning microwave microscope.

Our near-field scanning microwave microscope is based on an open-ended transmission line resonator of length L=1 m and characteristic impedance $Z_0$=50 Ω.[10,11] One end of the transmission line is connected to a fine open-ended coaxial probe positioned at a distance h above the sample, while the other end is weakly coupled through capacitance $C_d$ to a directional coupler and a frequency modulated microwave source $V_s$ (see Fig. 1). A frequency-following circuit is used to monitor the resonant frequency of the transmission line.[11] In this technique, the microwave source is frequency modulated by ±3 MHz at a rate of $f_{FM}$ = 3 kHz. The voltage reflected from the resonant cable is monitored by a diode at the output of the directional coupler. The diode output is fed to a lock-in amplifier, which is referenced at the oscillator frequency $f_{FM}$. The output voltage of the lock-in is integrated and the resulting voltage $V_{FM}$ is utilized to keep the source locked to a specific standing wave resonance $f_n$ of the transmission line. In this configuration, $V_{FM}$ is directly proportional to changes in $f_n$, caused by changes in coupling to the sample. In practice, we record $V_{FM}$ as a function of the position of the sample beneath the probe, thereby constructing a map of frequency shift versus position.

The interaction between the probe and a metallic sample can be represented by a capacitance $C_x$ between the sample and the inner conductor of the probe, analogous to the mechanism in scanning capacitance microscopy.[8-10] For a highly conducting sample and a small probe-sample capacitance $C_x$, i.e., $C_x \ll L/cZ_0$, the reflected voltage $V_r$ at the output of the directional coupler can be written as

$$|V_r| = \left| \left( \frac{Z_\alpha - Z_0}{Z_\alpha + Z_0} \right) P(1-P) V_s \right| \quad (1)$$

where

$$Z_\alpha = \frac{1}{i\omega C_d} + \frac{Z_0(\cosh(\gamma L) + i\omega C_x Z_0 \sinh(\gamma L))}{\sinh(\gamma L) + i\omega C_x Z_0 \cosh(\gamma L)} \quad (2)$$

and where c is the speed of light, $\omega$ is the angular frequency of the source, $\gamma$ is the propagation constant for the transmission line, and P is the coupling factor of the directional coupler. A plot of Eq. (1) versus frequency would show a series of dips corresponding to enhanced absorption at the resonant frequencies. For $C_x=0$ and $C_d=0$, the resonant frequencies are $f_n = nc/(2\sqrt{\varepsilon_r}L)$, for n=1,2, ..., where $\varepsilon_r$ is the dielectric constant of the transmission line; for our system a resonance occurs every 125 MHz. For small $C_x$, the n-th resonant frequency changes by:[8]

$$\Delta f_n \approx -f_n \frac{cC_x Z_0}{L} \quad (3)$$

Since $C_x$ depends on the distance between the inner conductor and the sample, equation (3) implies that $\Delta f$ can be used to determine the topography of the sample. When the sample is very close to the probe one expects that $C_x \approx A\varepsilon_0/h$ so that $\Delta f \sim -f_n c\varepsilon_0 A Z_0/hL$, where A is the area of the center conductor of the probe. On the other hand, when the sample is far from the probe, one must resort to numerical simulation to find $C_x(h)$. To find $C_x(h)$ for the coaxial probe geometry, we solved Poisson's equation using a finite difference method on a 200 by 150 cell grid. The solid line in Fig. 2 shows the corresponding calculated frequency shift $-\Delta f$ as a function of h. As expected we observe that as h approaches zero, the capacitance $C_x$ varies as $1/h$. For $h > a$, the frequency changes more slowly with height. For comparison, the points in Fig. 2 show data measured above a flat metallic surface at a frequency of $f_n$=9.572 GHz using a probe with an inner conductor diameter of 480 µm. Note that as the probe approaches the sample, the resonance frequency decreases ($\Delta f$ becomes more negative), corresponding to an increase in probe-sample capacitance $C_x$. The overall experimental behavior is qualitatively well described by the numerical simulation, including the weak frequency shifts observed at large separation. For convenience, we parameterize our measured calibration curves with an empirical function to easily transform measured frequency shifts to absolute heights.

To investigate the topographic imaging capabilities of the system, we imaged an uneven, metallic sample: a 1997 U.S. quarter dollar coin. First, the frequency shift was recorded as a function of position over the entire sample. Figure 3(a) shows a frequency shift image of the coin taken at 9.572 GHz using the 480 µm probe with a probe-sample separation of 30 µm at the closest point (at the left rim of the coin). The darker colored regions in Fig. 3(a) indicate a smaller distance (larger frequency shift) from the probe than the lighter regions (smaller frequency shift). We next measured the frequency shift versus height at a fixed position above a flat part of the sample (see Fig. 2) and determined the transfer function h($\Delta$f). Using h($\Delta$f), we then transformed the frequency shift image into a topographic surface plot (see Fig. 3(b)).

To determine the absolute accuracy of the height in the topographic image, we optically measured the height of several prominent features on the coin. In Fig. 4, the solid line represents a plot of height versus position as determined by the microwave microscope for a slice taken through the center of the coin (see arrows in Fig. 3a) while the open circles show the optically measured points. Corresponding points in the two data sets typically differ by about 13 µm. The accuracy at large separation is limited by the drift of the source frequency, and can be optimized by measuring the calibration curve immediately after the scan.[12]

To determine the height precision, the scan data at the left rim in Fig. 4 was fit to a parabola (over a range of x = ± 500 µm) and the standard deviation was found to be 55 nm for a probe-sample separation of 30 µm. Similarly, for the flat area on the table just to the right of the coin, a linear fit was performed to the data (over a range of x = ± 500 µm) and yielded a standard deviation of 40 µm at a probe-sample separation of 1.75 mm. In our system, the predominant source of noise is fluctuations $\delta(\Delta f)$ in the source frequency. In this case, the smallest detectable change in height is expected to be

$$\Delta h = \left.\frac{\partial h}{\partial \Delta f}\right|_h \delta(\Delta f) \quad (4)$$

For a source frequency fluctuation of 10 kHz and a value for the height transfer function of 41 μm/MHz we find that the theoretical height sensitivity at a probe-sample separation of 30 μm is approximately 41 nm, in reasonable agreement with the measured value.

In summary, we have used an open-ended resonant near-field scanning microwave microscope, to obtain quantitative topographic images of a metallic surface. As a result, we can quantitatively account for the contribution of topography to frequency shift images. Our technique allows for a height resolution of 55 nm for a 30 μm probe-sample separation and about 40 μm at a separation of 1.75 mm. The technique is simple and should be readily extendible to non-metallic samples, smaller probes, and closer or further separations.

We acknowledge support from the National Science Foundation NSF-MRSEC grant No. DMR-9632521, NSF grant No. ECS-9632811, and from the Maryland Center for Superconductivity Research.

12) A source frequency drift effectively translates the calibration curve along the $-\Delta f$-axis, and can be corrected by checking the frequency shift at a known height. In this case we used the probe-table separation of 1.75 mm.

Figures

Fig. 1. Schematic of the experimental setup and an uneven, uniformly conducting sample. Variation in probe-sample separation h produces a shift of the coaxial transmission line resonant frequency which is tracked by the frequency-following circuit.

Fig. 2. Calibration curve of height h versus frequency shift $-\Delta f$ taken above a flat section of a 1997 U.S. quarter dollar coin. The solid line represents a numerical simulation, whereas the dashed line shows $\Delta f \sim 1/h$, and the data points show measured values.

Fig. 3. (a) Frequency shift image of U.S. quarter dollar coin taken with a 480 µm diameter center conductor coaxial probe at 9.572 GHz. (b) Topographic surface plot of the coin.

Fig. 4. Height measured by an optical microscope (circles), compared with the topographic microwave data (solid line) obtained along line cut indicated by arrows in Fig. 3 (a).

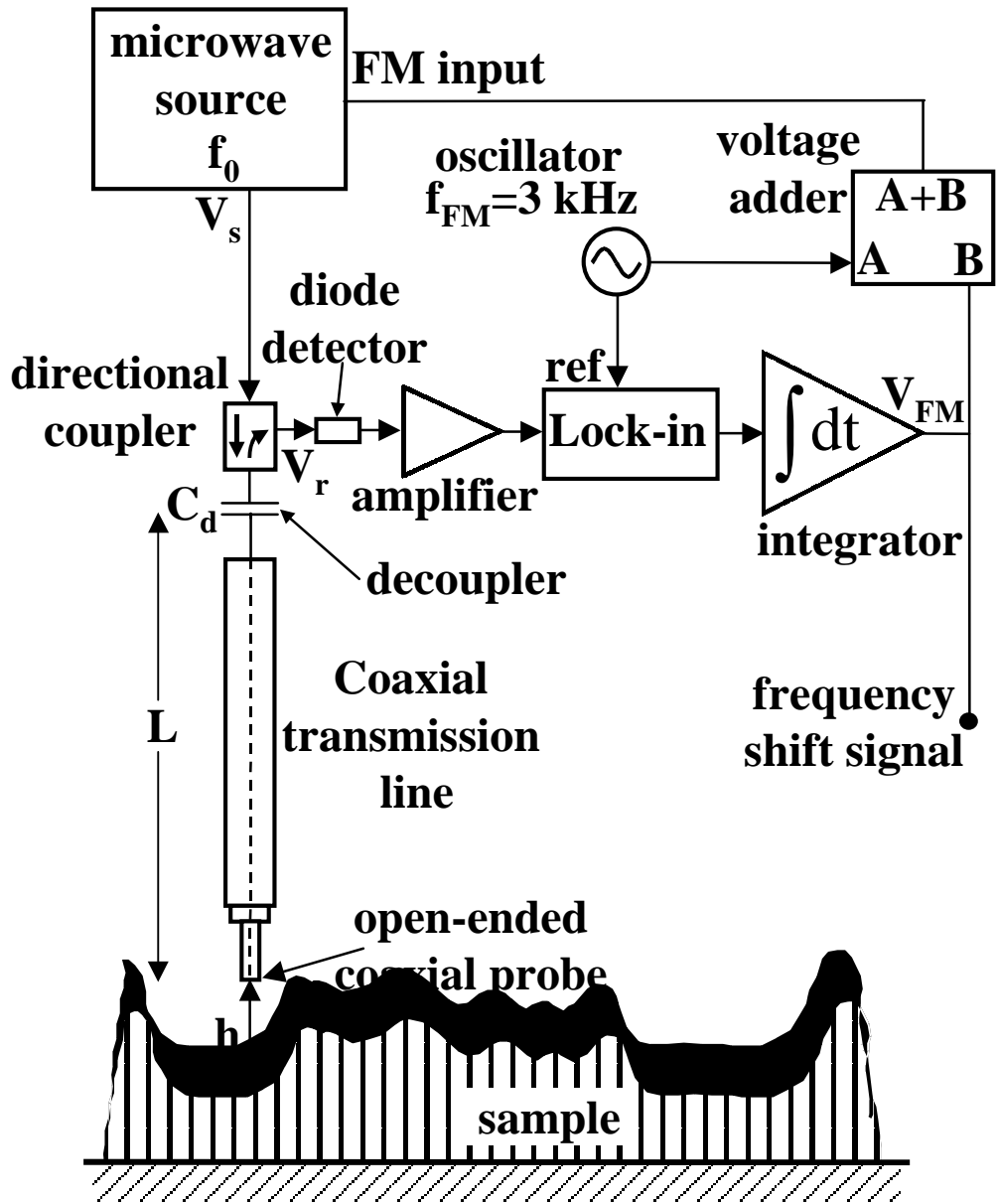

Figure 1

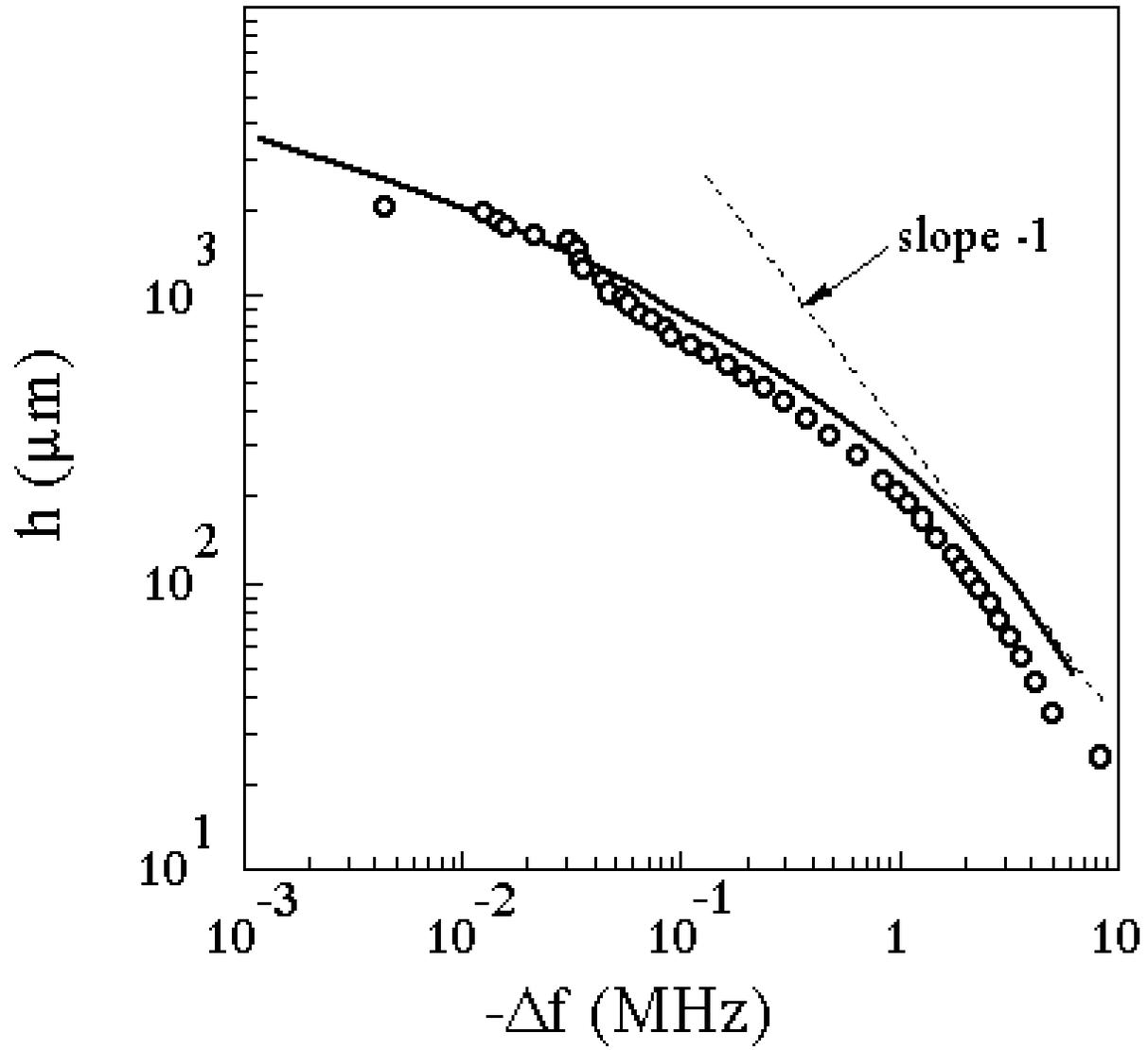

Figure 2

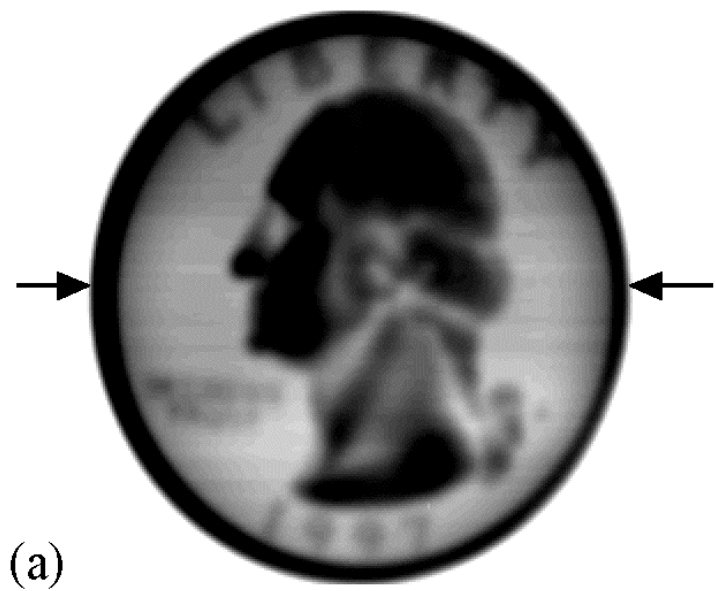

(a)

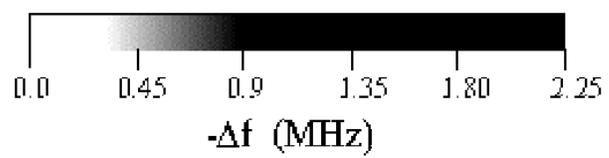

$-\Delta f$ (MHz)

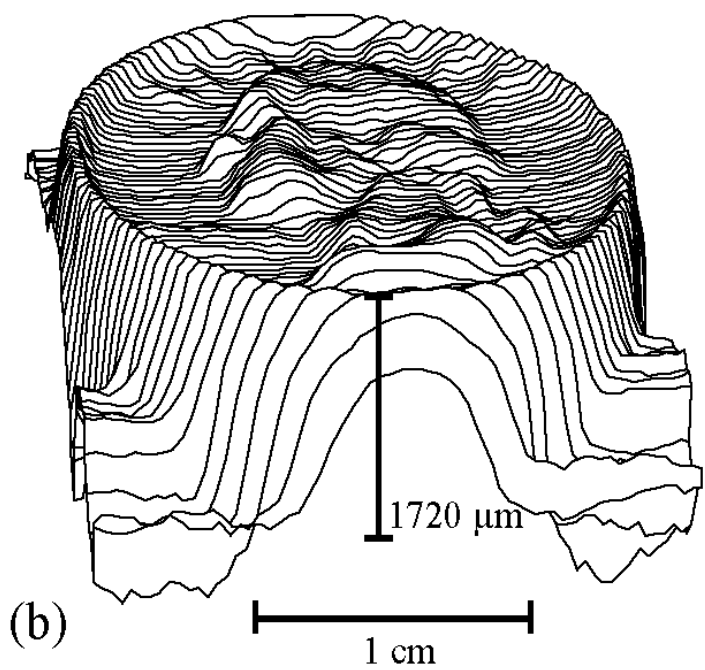

(b) 1720 μm

1 cm

Figure 3

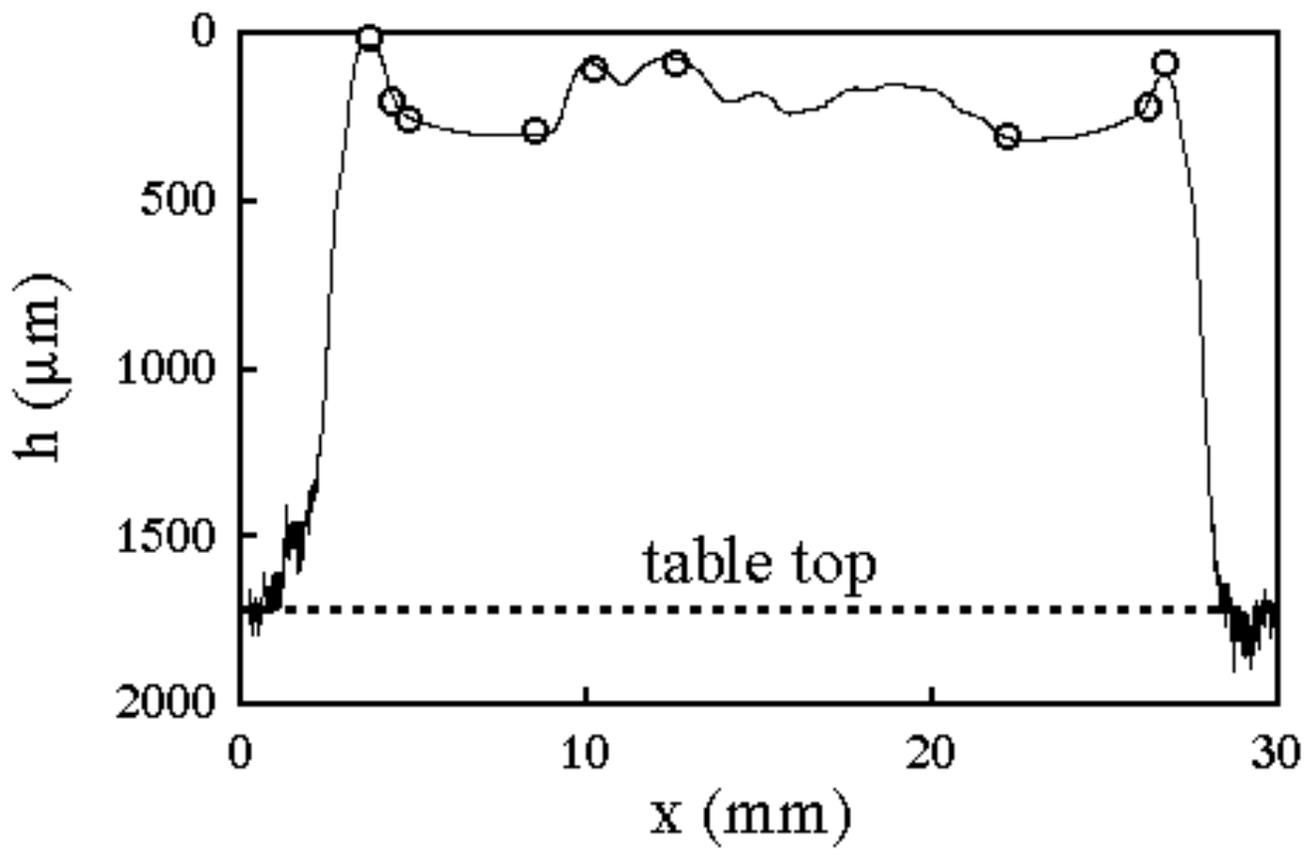

Figure 4